\documentclass{pcam_au}

\ArticleID{PCAM001} 

\usepackage{amsmath}

\def\ri{\mathrm{i}}

\def\sech{\mathop{\mathrm{sech}}}

\def\cn{\mathop{\mathrm{cn}}}

\begin{document}

\title{The Korteweg--de Vries Equation}

\author{W.\ A.\ Hereman}

\affiliation{Colorado School of Mines}

\maketitle

\section{Historical Perspective}

In 1895, Diederik Korteweg (1848--1941) and Gustav de Vries (1866--1934) 
derived a partial differential equation (PDE) that models the 
``great wave of translation" that naval engineer John Scott Russell 
had observed in the Union Canal in 1834.

Assuming that the wave propagates in the $X$-direction, 
the evolution of the surface elevation 
$\eta(X,T)$ above the undisturbed water depth $h$ at time $T$ 
can be modeled by the Korteweg--de Vries (KdV) equation,
\begin{align}
&
\frac{\partial \eta}{\partial T} 
+\sqrt{gh}\, \frac{\partial \eta}{\partial X}
+ \frac{3}{2}\frac{\sqrt{gh}}{h} \eta \frac{\partial \eta}{\partial X} 
\nonumber \\
&
\quad\mbox{}
+ \frac{1}{2} h^2 \sqrt{gh} 
\left(\frac{1}{3} - \frac{{\mathcal T}}{\rho g h^2} \right) 
\frac{\partial^3 \eta}{\partial X^3} = 0,  
\label{kdvoriginal}
\end{align}
where $g$ is the gravitational acceleration, 
$\rho$ is the density, and ${\mathcal T}$ is the surface tension.
The dimensionless parameter ${\mathcal T}/\rho g h^2,$ called the 
{\it Bond number}, measures the relative strength of surface tension and 
the gravitational force.
Equation (\ref{kdvoriginal}) is valid for long waves (or shallow water) 
of relatively small amplitude, $| \eta |/h \ll 1$.

In dimensionless variables, (\ref{kdvoriginal}) can be written as
\begin{equation}
\label{kdvdimless}
    u_t + \alpha u u_x + u_{xxx} = 0, 
\end{equation}
where subscripts denote partial derivatives.
The term $\sqrt{gh} \, \eta_X$ in (\ref{kdvoriginal}) has been removed by 
an elementary transformation.
Conversely, a linear term in $u_x$ can be added to (\ref{kdvdimless}).
The parameter $\alpha$ can be scaled to any real number. 
Commonly used values are $\alpha = \pm 1$ and $\alpha = \pm 6$. 

The term $u_t$ describes the time evolution of the wave. 
Therefore, (\ref{kdvdimless}) is called an {\it evolution} equation. 
The nonlinear term $\alpha u u_x$ accounts for steepening of the wave.
The linear dispersive term $u_{xxx}$ describes spreading of the wave. 

The KdV equation had already appeared in seminal work on 
water waves published by Joseph Boussinesq about 20 years earlier.

\section{Solitary wave and periodic solution}

The balance of the steepening and spreading effects gives rises to a stable 
solitary wave,  
\begin{equation}
\label{kdvsolitary}
   u(x,t) = \frac{\omega -4 k^3}{\alpha k} 
   + \frac{12 k^2}{\alpha} {\sech}^2 (k x - \omega t + \delta),
\end{equation}
where the wave number $k$, the angular frequency $\omega,$ and phase 
$\delta$ are arbitrary constants. 
Requiring that $\lim_{x \to \pm \infty} u(x,t) = 0$ for all $t$ leads to 
$\omega = 4 k^3$,
in which case (\ref{kdvsolitary}) reduces to 
\begin{equation}
\label{kdvsolitarysimpler}
    u(x,t) = 12 (k^2/\alpha)\, {\sech}^2 (k x - 4 k^3 t + \delta). 
\end{equation}
This hump-shape solitary wave of finite amplitude $12 k^2/\alpha$ travels 
to the right at constant phase speed $v = \omega/k = 4 k^2.$ 
It models the ``great wave of translation" that traveled without change of 
shape over a fairly long distance as observed by Scott Russell.

As shown by Korteweg and de Vries, equation (\ref{kdvdimless}) also has a 
periodic solution,
\begin{align}
  u(x,t) & = (\omega - 4 k^3 (2 m - 1))/(\alpha k) \nonumber\\
      &\quad\mbox{}
      + 12 (k^2 /\alpha) m \,{\cn}^2 (k x -\omega t + \delta;\, m),
\label{kdvcnoidal}
\end{align}
which they called the {\it cnoidal wave} solution for it involves
Jacobi's elliptic cosine function, $\cn,$ with modulus $m$,
$0<m<1$. In the limit $m\to 1$,
 $\cn(\xi; m) \to  \sech \xi$ and 
(\ref{kdvcnoidal}) reduces to~(\ref{kdvsolitary}).

\section{Modern Developments}

The solitary wave was for many years
considered an unimportant curiosity in the field of 
nonlinear waves.
That changed in 1965, when Zabusky and Kruskal realized that the KdV 
equation arises as the continuum limit of a one-dimensional anharmonic 
lattice used by Fermi, Pasta, and Ulam in 1955 to investigate how energy is 
distributed among the many possible oscillations in the lattice. 
Since taller solitary waves travel faster than shorter ones, Zabusky and 
Kruskal simulated the collision of two waves in a nonlinear crystal lattice 
and observed that each retains its shape and speed after collision.
Interacting solitary waves merely experience a phase shift, advancing the 
faster and retarding the slower wave.
In analogy with colliding particles, they coined the word ``solitons" 
to describe these elastically colliding waves.

To model water waves that are weakly non\-linear, weakly dispersive, and 
weakly two-di\-men\-sional with all three effects being comparable, 
 Kadomtsev and Petviashvili (KP) derived a two-dimensional version 
of (\ref{kdvdimless}) in 1970
\begin{equation}
\label{kpdimless}
(u_t + 6 u u_x + u_{xxx})_x + 3 \sigma^2 u_{yy} = 0, 
\end{equation}
with $\sigma^2 = \pm 1$ and where the $y$-axis is perpendicular to the 
direction of propagation of the wave (along the $x$-axis).

The KdV and KP equations and the nonlinear Schr\"odinger (NLS) equation,
\begin{equation}
\label{nlsdimless}
\ri u_t + u_{xx} + \kappa |u|^2 u = 0, 
\end{equation}
where $\kappa$ is a constant and $u(x,t)$ is a complex-valued function, 
are famous examples of so-called completely integrable nonlinear PDEs 
which can be solved with the inverse scattering transform (IST), 
a nonlinear analog of the Fourier transform.  

The IST is not applied to (\ref{kdvdimless}) directly but to an auxiliary 
system of linear PDEs, 
\begin{align}
\label{kdvlax1}
\psi_{xx} + (\lambda + \frac{\alpha}{6} u) \psi &=0, \\
\label{kdvlax2}
\psi_t + \frac{\alpha}{2} u_x \psi + \alpha u \psi_x + 4 \psi_{xxx} 
&=0,
\end{align}
which is called the {\it Lax pair\/} for the KdV equation.
Equation (\ref{kdvlax1}) is a linear Schr\"odinger equation for an 
eigenfunction $\psi$, a constant eigenvalue $\lambda$, and 
a potential $(- \alpha u)/6$.
Equation (\ref{kdvlax2}) governs the time evolution of $\psi$.
The two equations are compatible, i.e., $\psi_{xxt}=\psi_{txx}$, 
if and only if $u(x,t)$ satisfies (\ref{kdvdimless}). 
For given $u(x,0)$ decaying sufficiently fast as $|x| \to \infty$, 
the IST solves (\ref{kdvlax1}) and (\ref{kdvlax2}) and finally 
determines $u(x,t)$.

\section{Properties and Applications}

Scientists remain intrigued by the rich mathematical structure of completely 
integrable nonlinear PDEs that can be written as infinite-dimensional 
bi-Hamiltonian systems. 
Completely integrable nonlinear PDEs have remarkable features, 
such as a Lax pair, a Hirota bilinear form, B\"acklund transformations, 
and the Painlev\'e property. 
They have an infinite number of conserved quantities, infinitely many 
higher-order symmetries, and an infinite number of soliton solutions.

Apart from shallow water waves, the KdV equation is ubiquitous in applied 
science. 
It describes, for example, ion-acoustic waves in a plasma, 
elastic waves in a rod, and internal waves in the atmosphere or ocean. 
The KP equation models, e.g., water waves, acoustic waves, and magneto\-elastic
waves in anti-ferro\-magnetic materials.
The NLS equation describes weakly nonlinear and dispersive wave packets in 
physical systems, e.g., light pulses in optical fibers, surface waves in 
deep water, Langmuir waves in a plasma, and high-frequency vibrations on a 
crystal lattice. 
Equation (\ref{nlsdimless}) with an extra linear term $V(x) u$ to account for 
the external potential $V(x)$ also arises in the study of 
Bose--Einstein condensates, where it is referred to as the time-dependent 
Gross--Pitaevskii equation.

\section*{Further Reading}

\noindent
Ablowitz M.\ J.\ and Clarkson P.\ A.,
Solitons, Nonlinear Evolution Equations and Inverse Scattering, 
Cambridge University Press, Cambridge, U.K., 1991.

\smallskip

\noindent
Kasman A., 
Glimpses of Soliton Theory, AMS, Providence, RI, 2010.

\smallskip

\noindent
Osborne A.\ R.,
Nonlinear Ocean Waves and the Inverse Scattering Transform,
Academic Press, Burlington, MA, 2010.

\vfill
\noindent
{\bf Revised June 14, 2013. \\
 Edited by PAM, 30 July 2013\\
Comments by NJH, 4 August 2013\\
 Edited by PAM, 18 August 2013}\\
 {\tt Hereman-KdVr.tex}

\end{document}